\documentstyle[twocolumn,aps]{revtex}

\begin{document}
\draft
\title{A protocol for secure and deterministic quantum key expansion}
\author{Xiang-Bin Wang\thanks{Email address: wang@qci.jst.go.jp}\\
IMAI Quantum Computation and Information Project,
ERATO, JST,
Daini Hongo White Bldg. 201, \\5-28-3, Hongo, Bunkyo,
Tokyo 133-0033, Japan}
\maketitle
\begin{abstract} 
 In all existing protocols of private communication with encryption and 
decryption, the pre-shared key can be used for only {\it one time}. 
 We give a deterministic quantum key expansion 
protocol where the pre-shared key can be recycled. Our protocol
costs less qubits and almost zero classical communication. 
Since the bit values of the expanded key is deterministic,
this protocol can also be used for direct communication.
 Our protocol includes the authentication steps therefore we 
don't worry about the case that
Alice and Bob are completely isolated.
\end{abstract}
%%%%%%%%%%%%%%%%%%%%%%%%%%%%%%%%%%%%
\section{introduction}
%%%%%%%%%%%%%%%%%%%%%%%%%%%%%%%%%%%%
Information processing with quantum systems enables us to do novel tasks
which seem to be impossible with its classical counterpart
\cite{wies,shor,bene}.
Among all of the non-trivial quantum algorithms,
quantum key distribution (QKD)
\cite{bene,eker,ben2,ben3,brus,gisi,maye}
is one of the most important and interesting
quantum information processing due to its relative low technical overhead:
the only thing required there is quantum states preparing, transmission and measurement. 
It needs neither quantum memory nor collective quantum operation such
as the controlled-NOT (CNOT) gate.
Therefore, QKD will be the first practical quantum information
 processor \cite{gisi}.  
QKD makes it possible for two remote parties, 
Alice and Bob to make unconditionally 
secure communications: they first build up a secure shared key and then
use this key as the one-time-pad to send the private message.
However, in the standard BB84\cite{bene} protocol, at least half of the transmitted qubits
are discarded due to the mismatch of preparation bases and measurement bases to
the qubits. Also, the standard BB84 protocol does not include authentication.
This makes it insecure in the case that Alice and Bob are completely isolated:
Eavesdropper (Eve) may intercept all classical information and quantum information and 
the actual case there 
is that each of Alice and Bob are doing QKD with Eve separately.

In this Letter, we shall give an efficient protocol to expand the key deterministically
or make direct communication, with
authentication being included. Our protocol 
has the advantage of lower cost in both classical communication and quantum states transmission.
 Our protocol includes the authentication steps.
The pre-shared key can be recycled in our protocol. 

The requirement of pre-sharing a secret string is not a serious drawback
of our protocol. In the case authentication is required for security,
all protocols need a pre-shared secret string; 
in the case that authentication is thought to be unnecessary,  
our protocol  need not pre-share anything initially:
they may first use any standard QKD protocol to 
generate a secret random string and then use this string as the pre-shared string.

The the initial version of QKD protocol\cite{bene} proposed by Bennett and Brassard 
is fully efficient by delaying the measurement.
This delay requires  the quantum memories which
are very difficult technique.
 Another method is to assign
 significantly different probabilities to the different bases
\cite{lo2}. Although unconditional security of the scheme is
given \cite{lo2}, it has a disadvantage that a larger number of
 key must be generated at one time.
Roughly speaking, with the  bases mismatch rate being set to  $\epsilon$,
 the number of qubits it
needs to generate at one time is $\epsilon^{-2}$ times of that of the
standard BB84\cite{bene}. In a
 recently proposed QKD protocol without public announcement of basis
 (PAB) \cite{hwa2,wang}, there is no measurement mismatch. 
However, the protocol in its present form has the disadvantage
that one must make many batches of keys before any batch is used to
encrypt and transmit classical message. Note that 
  they must abort the preshared secret string after the key expansion.
To really have an advantage in the efficiency, one should generate
as many secret bits as possible at one time, by that protocol. 
Blindly generating too many secret bits at one time means a higher cost:
First, the complexity of decoding the error correction code rises rapidly
with the size of the code. Second, the quantum channel could be expensive.
In practice, it could be the case that we don't know how many secret bits are 
needed in the future communication. 
For example, a detective is sent to his enemy country Duba from the country VSA. 
He is scheduled
to work in Duba for only one month and then come back to the headquarter
in VSA. The so called secret bits will be useless after that month.

The existing protocols for quantum direct communication can save some 
cost of classical communications. Unfortunately, they are either
insecure\cite{cai,hoff,woj} or only quasisecure\cite{guo}.
Moreover, all of them require quantum memory. 

So far, it seems that our protocol is the unique one which has the advantage of 
lower cost of both quantum states transmission and classical communication
while still holding the unconditional security.
\section{ our protocols and security proof}
We shall use the reduction technique. We first reduce the classical protocol
to quantum protocol (the one uses perfect entangled pairs and quantum memories), and then reduce
the quantum protocol back to classical protocol (the one without any entangled
 pair or quantum memory).
We start with a trivial scheme, {\it Protocol 1}.\\
{\it Protocol 1, Classical protocol}\\
Alice and Bob share a secret key, i.e., $g$-bit random string, $G$. Alice wants to
send an $N$-bit classical binary string $s$ to Bob, $g>N$. She chooses 
 first $N$ bits from $G$ and denotes 
this substring as $b$. She prepares an $N$-qubit string $q$
which is in the 
quantum state $|b\oplus s\rangle$, and sends these $N$ qubits to Bob. Here $\oplus$
is the summation modulo 2. Suppose the values of the
$i$th element in string $b$ and $s$ are $b_i$ and $s_i$, respectively,
 given any value of $b_i\oplus s_i$, she just prepares the $i$th quantum state
 $|b_i\oplus s_i\rangle$ accordingly. All qubit states in $q$ are  prepared
in $Z$ basis. 
Bob measures each of qubits in $Z$ basis
and obtain an $N-$bit classical string, taking 
$\oplus$ operation of this string
and string $b$ he obtains the message string. Alice and Bob discards
string $b$. 

This is just 
classical private communication with one-time-pad. 
Obviously, the message string $s$ is perfectly
secure no matter how noisy the quantum channel is.
Though there are bit-flip errors in to the transmitted message, 
there is no information leakage. 
 In this protocol, the
one-time-pad cannot be recycled. Since all qubits are prepared in $Z$ basis,
Eve in principle can have full information of $b\oplus s$ without disturbing
the quantum string $q$ at all.
 For the purpose of recycling the one-time-pad, we reduce it to our
{\it Protocol 2}, a quantum protocol. Latter on, we shall classicalize
     {\it Protocol 2}.\\
{\it Protocol 2:} Secure communication with recyclable quantum one-time-pad.\\
Alice and Bob share $g$ pairs of (exponentially) perfect entangled pairs of 
$|\phi^+\rangle= \frac{1}{\sqrt 2}(|00\rangle+|11\rangle)$. For convenience
we shall call this pair state as EPR pair.
Alice wants to
send $N$-bit classical binary string $s$ to Bob. According to each individual
bit information, she prepares an $N-$qubit quantum state $|s\rangle$, all of
them being prepared in $Z$ basis. 
She chooses her halves of first $N$ 
pairs from $g$ pairs and number them from 1 to $N$. We denote these $N$ pairs
by $E$, Alice's halves of $E$ by $E_A$, Bob's halves of $E$ as $E_B$.
 To each of the $i$th qubit in $|s\rangle$
and $i$th qubit in $E_A$, she takes CNOT operation with
the $i$th qubit in $E_A$ being the controlled
qubit and the $i$th qubit in $|s\rangle$ as the target
qubit. $i$ runs from 1 to $N$. She sends those $N$ target qubits to Bob.
Bob takes a CNOT operation to each of the $i$th received qubit and the 
$i$th qubit of $E_B$, with the received qubit being the 
target qubit and the qubit in $E_B$ being controlled qubit.
Bob takes a measurement in $Z$ basis to each of the target qubit and obtain
a classical string. He uses this string as the message from Alice.\\
The message $s$ in this protocol is as secure as that in {\it Protocol 1}.
\\
{\it Proof.}
Imagine the case that Alice measures each qubits in $E_A$ in $Z$ basis
in the beginning, then {\it protocol 2} is identical to {\it Protocol 1}.
However, no one except Alice knows whether she has taken the measurement.
Therefore she can choose not to measure her halves of entangled pairs.
This is just {\it Protocol 2}. In {\it Protocol 2}, $N$ EPR pairs have 
been used as a quantum shared key, however, we don't have to
discard them after the message $s$ has been decrypted. 
Instead, Alice and Bob may do purification to those $N$ pairs, given the 
information of 
bit-flip rate and phase-flip rate. After the purification, the outcome pairs
can be re-used as (almost) perfect entangled pairs. So the next question is on how 
to do the purification efficiently. The bit-flip rate is defined as the percentage of pairs
which have been changed into state 
$|\psi^+\rangle=\frac{1}{\sqrt 2}(|01\rangle +|10\rangle)$ or state 
$|\psi^-\rangle=\frac{1}{\sqrt 2}(|01\rangle -|10\rangle)$; 
phase-flip rate is defined as the percentage of pairs
which have been changed into state 
$|\phi^-\rangle=\frac{1}{\sqrt 2}(|00\rangle -|11\rangle)$ or state 
$|\psi^-\rangle$. Or mathematically, 
if we consider the Pauli channel consisting of the following operations:
\begin{eqnarray}
\sigma_x= ( \begin{array}{cc} 0 & 1  \\
                                       1 & 0
             \end{array} ),
 \sigma_y= ( \begin{array}{cc} 0 & -i  \\
                                            i &  0
             \end{array} ),
\sigma_z= ( \begin{array}{cc} 1 & 0  \\
                                         0 & -1
             \end{array} )\end{eqnarray}
    the channel operation $\sigma_x$ or $\sigma_y$ causes a bit-flip,
the channel operation $\sigma_y$ or $\sigma_z$ will cause a phase-flip.
One direct way to know the bit-flip rate and phase-flip rate is to let
Alice and Bob randomly take some samples of those pairs and then measure
the samples in $Z$ ( $\{|0\rangle,|1\rangle\}$)or in $X$ 
( $\{|\pm\rangle=\frac{1}{2}(|0\rangle\pm |1\rangle\})$)basis in each side, 
and obtain the statistical
values of those flip rates for the remained pairs. However, 
in testing the phase-flip rates with samples of those used EPR pairs,
the corresponding message bits must be discarded because once the bit values
of EPR pairs are announced, Eve has a way to attack encrypted message bits.
Moreover, we want to reduce
the protocol back to classical protocol therefore we don't directly sample the
entangled pairs. We can have a better way for the error test. 
Consider the initial state of an entangled pair and the 
the quantum state of message bit $|\chi_A\rangle$,
\begin{eqnarray}
|h_0\rangle = |\phi^+\rangle\otimes |\chi_A\rangle.\label{initial}
\end{eqnarray}
In the most general case $|\chi_A\rangle=\alpha |0\rangle +\beta |1\rangle$ and $|\alpha|^2+|\beta|^2=1$.
In our {\it protocol 2}, there will be Alice's  CNOT operation, transmission and Bob's CNOT operation to the message qubit. 
In transmission,  the encrypted quantum state of message bit
could bear a flipping error of  $\sigma_x,\sigma_z$ or $\sigma_y$.
It is easy to
see that, after Bob's CNOT operation, $\sigma_x$ error of transmission channel
will cause a $\sigma_x$ error to 
the the message qubit only,  $\sigma_z$ error of 
transmission channel will cause a $\sigma_z$ error to 
the EPR pair and $\sigma_z$ error to the message qubit, while $\sigma_y$ error 
of transmission channel will cause a $\sigma_z$ error to the EPR pair $and$ a $\sigma_y$ error
to the message qubit. That is to say, the final state will be
\begin{eqnarray}
|h_f\rangle=|\phi^+\rangle\otimes (\sigma_x |\chi_A\rangle)
\end{eqnarray} 
given a $\sigma_x$ flip to the encrypted message qubit in transmission;
\begin{eqnarray}
|h_f\rangle=(\sigma_z|\phi^+\rangle)\otimes (\sigma_z |\chi_A\rangle).\label{phase}
\end{eqnarray} 
given a $\sigma_z$ flip to the encrypted qubit in transmission;
and 
\begin{eqnarray}
|h_f\rangle=(\sigma_z|\phi^+\rangle)\otimes (\sigma_y|\chi_A\rangle)
\end{eqnarray} 
given a $\sigma_y$ flip to the encrypted qubit in transmission. 
We now show eq.(\ref{phase}). The other two equations can be shown in a similar way.
Consider the initial state  defined by eq.(\ref{initial}). After the CNOT operation done by Alice, the state is changed
to
\begin{eqnarray}
|h_0'\rangle=\frac{1}{\sqrt 2}|00\rangle\otimes |\chi_A\rangle +\frac{1}{\sqrt 2}|11\rangle \otimes(\alpha |1\rangle+
\beta|0\rangle). 
\end{eqnarray}
Suppose there is a phase-flip to the encrypted qubit during the transmission, the total state is then changed to
\begin{eqnarray}
|h_0''\rangle = \frac{1}{\sqrt 2}|00\rangle\otimes (\alpha |0\rangle-\beta|1\rangle) 
+\frac{1}{\sqrt 2}|11\rangle \otimes(-\alpha |1\rangle+
\beta|0\rangle).
\end{eqnarray} 
After Bob take the CNOT operation, the final state is changed to
\begin{eqnarray}
|h_f\rangle=|\phi^-\rangle\otimes |\chi_A\rangle= (\sigma_z |\phi^+\rangle)\otimes (\sigma_z |\chi_A\rangle).
\end{eqnarray} 
This completes the proof. Although there could be phase-flips to the transmitted qubits, as we have shown already,
in principle, there is no information leakage of the original message. Threfore we disregard those phase-flips to the
message qubits. 
Note that the model of Pauli channel and classical statistics work perfectly here\cite{sho2,lo3,wangs}, given $arbitrary$
channel noise, including any type of collective noise.
Therefore if we know the bit-flip rate and phase-flip rate of the channel, we can deduce exactly
the flipping rate of those used EPR pairs. 
Therefore we can simply
mix some of qubits (test qubits) in transmitting the message qubits. 
We don't do any CNOT operations (quantum encryption or decryption) 
to those test
qubits. Half of the test qubits should be prepared in $X$ basis and half of the test qubits should be prepared in
$Z$ basis. All of the test qubits should be mixed randomly with the message qubits. Bob needs to know the measurement
bases of each qubits so as not to destroy any message qubits. 
Bob also needs to know which qubits are for testing
and the original state of each test qubits so as to see the 
flip-rates of transmission.  Therefore,  besides $N$ EPR pairs,
they must also share a classical
 string $b'$ for the information of bases, positions
and bit values of each test qubits. 
Suppose after reading the test qubits, Bob finds the error rate to
those test qubits in $X$ bases is
$t_0$. Then they may safely assume $(t_0+\delta)N$ phase-flips to the
used EPR pair.
$\delta$ is a very small number. 
The probability that the phase-flip rate of those used EPR pairs 
is larger than $t_0+\delta$
is exponentially small. As we have shown earlier, there is not bit-flip error to the used entangled pairs.
Therefore they may purify the used pairs by the standard purification protocol\cite{sho2,ben4} which costs
only $N\cdot H(t_0+\delta)$ pairs, 
$H(x)= -x\log_2 x -(1-x)\log_2(x)$.

Since their purpose is to re-use those pairs securely for private 
communication in the future instead of
really reproducing the  perfect EPR pairs, 
they need not really complete the full procedure of the purification.
Instead, as it has been shown in Ref\cite{sho2}, 
except Alice herself, no one knows it if she measures 
all EPR pairs in $Z$ bases
in the begining of the protocol. 
Therefore the CSS code can be 
classicalized\cite{sho2} if the purpose is for security of
private communication instead of the real entanglement purification
. Consequently, 
the initially shared EPR pairs before running the protocol
can be replaced by a classical random string and after they run 
the protocol they recycle the random string
by a classical Hamming code with the phase-error rate input being $t_0+\delta$.
{\it Protocol 3} can help them to do quantum key expansion efficiently, without any quantum memory or 
entanglement resource:
\\
1. Alice and Bob pre-share a secret classical random string $G$.
They are sure that the bit-flip rate and phase-flip rate
 of the $physical$ channel
are less than $t_x-\delta$ and $t_z-\delta$, respectively. 
(In quantum cryptography, the knowledge of flipping
rates of physical channel does not guarantee the security in any sense.)
They choose two Hamming code $C_x$ and $C_z$ which can correct
$(t_x+\delta)M$ bits and $(t_z+\delta)M$ bits of error, respectively.
We suppose $t_x+\delta < 11\%$ and $t_z+\delta < 11\%$. 
2. Alice plans to send $N$ deterministic bits, string $s$ to Bob.
 Alice and Bob 
take an $M-$bit substring $b$, an $M'-$bit substring $b'$, a 200-bit
substring $c$ and a 200-bit substring $d$ from $G$, from left to right. 
Here $M= \frac{N}{1-H(t+\delta)}$.
3. Alice expands the message string $s$ to $S$
by Hamming code $C_x$. Obviously, there are $M$ bits in the expanded
string $S$. 
She encrypts the expanded string $S$ with string $b$, i.e., she prepares
an $M-$qubit quantum state $|S_q\rangle=|S\oplus b\rangle$ in $Z$ basis.
All these encrypted message qubits are placed in order. She also 
produces $rN=2k$ test qubits and mix them with those qubits in $|S_q\rangle$.
The position, bit value and preparation basis of each test qubits 
are determined by substring $b'$. This requires substring $b'$ including 
$M'==\left(\begin{array}{c}M+2k\\2k\end{array}\right)+4k$ bits.
 The bit values (0 or 1), position and bases ($X$ or
$Z$) of those test qubits must be totally random, since $b'$ is random.
After the mixing, she has a quantum sequence $q$ which contains
 $M+2k$ qubits.
4. Alice transmits sequence $q$ to Bob.
5. Bob reads $b'$. After receives sequence $q$ from Alice, 
he measures each of them in the correct bases. 
He then separates the test bits and message bits, 
with their original positions in each string being recovered.
Bob reads the test bits and check the error rate (authentication). 
If he finds the 
bit-flip rate $t_{x0}> t$ or 
phase-flip rate $t_{z0}> t$ on the test bits, he sends substring $c\oplus d$
to Alice by classical communication and abort the protocol with string
$c$ being deleted from $G$.
If he finds the bit-flip rate $t_{x0}\le t$ and 
phase-flip rate $t_{z0}\le t$ on the test bits, he sends substring $c$
to Alice by classical communication  and continues the protocol.
6. Bob deletes $c$ from $G$. 
He decrypts the encrypted expanded message string by $b$ and then
decodes it by Hamming code $C_x$ and obtains the message string. 
The probability that Bob's decoded string is not identical 
to the original message string $s$ is exponentially close to 0.
The key expansion part (or communication part) has been completed now. 
7. Alice reads the 200-bit classical message from Bob.
If it is not $c$, she aborts the protocol with string $c$ being deleted
from $G$.(This is also authentication.) If it is $c$, she deletes
substring $c$ from $G$ and  carries out 
the next step.
8. Alice and Bob replace $b$ by the coset of $b+C_z$ 
as the recycled string.\\
{\it Remark 1}.
Our cost of qubit-transmission is less than half of
that in BB84 protocol. Our cost of classical communication is almost
zero.
{\it Remark 2}. After the protocol, string $b'$ and $d$ can be re-used safely. 
In our protocol, even Alice announces $b',d$, Eve's information
about message $s$ is 0. Therefore the mutual information between 
$s$ and $\{b',d\}$ is
$I(s:\{b',d\})=0$. Therefore, if message $s$ is announced while $\{b',d\}$ 
is not announced, 
Eve's information about $\{b',d\}$ must be also 0. Consequently, Eve's
information to $\{b',d\}$ must be zero after the protocol.\\
{\it Remark 3.} If we want to reduce the number of pre-shared qubits, we can
use fewer test bits, i.e., reduce the value of $r$. In our protocol,
the total qubits needed is $r^{-1}$ times of that of BB84 protocol. To avoid
a too large key expansion at one time, we can choose to raise the value
of $\delta$, given a small $r$. 
\section{Existing protocols of direct communication with qubits are insecure.}
Our protocol cannot be replaced by 
any existing direct communication protocol\cite{bf,cai,hoff,woj,guo} with quantum states.
The insecurity of existing direct communication protocols have been pointed
out already for the case of noisy channel\cite{cai,hoff}. Here we show
that these protocols are not exponentially secure even with noiseless
quantum channel. We suppose that there are $m$ test qubits and $N$ message qubits.
Consider the best case that they find no error to 
the test bits. Even in such a case, the message is still polynomially insecure: Eve has
non-negligible probability to obtain a few bits information to
the message. For example, Eve just intercepts one qubit in transmission
and measures it in $Z$ basis ($\{|0\rangle,|1\rangle\}$)  and then resends it to Bob. 
Suppose the physical channel itself is noiseless. Obviously, There is 
a probability of $N/(N+m)$ that Bob finds no error to the test bits while Eve
has one bit information about the message. In particular, in certain
cases, 1 bit leakage of message is disastrous\cite{lo3}. 
Such type of
direct private communication is insecure {\it even with noiseless quantum channel},
since the zero error of test bits only guarantees less than $\delta$ errors of
the message bits, it does not guarantee zero phse-flip of the message bits.
{\it In principle, there is no way to verify zero phase-flip error of the untested 
bits by looking
at the test bits only.}
The insecurity of existing protocols 
is due to the lack of privacy amplification step, which is the main
issue of the security of private communication. 
One cannot directly append a privacy amplification
step here since this may change the message bits therefore destroy the message.
One of the non-trivial point of our protocol is that {\it the transmitted message bits
in our protocol is  unconditionally secure without any privacy amplification, no matter
how noisy the channel is}. There we only need to correct the $bit$-flip 
errors in
the message. This does not change the message itself.
\section{discussions}
Our protocol can obviously be used for both key expansion and direct communication.
In the security proof, we have used a pre-condition that Eve has zero 
information to the preshared string $G$. However, strictly speaking,
this condition does not hold in our real protocol. First, as we have argued 
that the pre-shared string can be generated by standard BB84 QKD protocol where
Eve's information to the shared key is exponentially close to 0 instead of
strict 0. Second, 
Eve's information to the recycled string is also exponentially close to 0 
rather than 0. Eve's exponentially small prior information is not a problem
to the security of classical private communication. However, here we have 
used quantum states to carry
the classical message, Eve may store her $quantum$ information about the 
pre-shared (or recycled) secret string and directly attacks 
the decoded message or the updated key finally. 
With the universality
of quantum compossibility\cite{comp}, we know that Eve's a exponentially small amount
of prior information about the pre-shared string or the recycled string will only cause an exponentially
small amount of information about the private message or the updated shared string.  Therefore our protocol is unconditionally secure in the real case that
Eve has exponentially small amount of information to the pre-shared key.

%%%%%%%%%%%%%%%%%%%%%%%%%%%%%%%%%%%
\acknowledgments
%%%%%%%%%%%%%%%%%%%%%%%%%%%%%%%%%%%
I am
very grateful to Prof. H. Imai for his long term supports. I thank D. Leung and H.-K. Lo for pointing
out ref\cite{comp}.

\end{document}